# Material-structure integrated design for ultra-broadband microwave metamaterial absorber

*Mengyue Peng[a], Faxiang Qin[a,*], Liping Zhou[a], Huijie Wei[a], Zihao Zhu[a], Xiaopeng Shen[b]*

[a] Institute for Composite Science Innovation (InCSI), School of Materials Science and Engineering, Zhejiang University, Hangzhou, 310027, China

[b] School of Material Science and Physics, China University of Mining and Technology, Xuzhou, 221116, China

**Abstract**

We propose herein a method of material-structure integrated design for broadband absorption of dielectric metamaterial, which is achieved by combination of genetic algorithm and simulation platform. A multi-layered metamaterial absorber with an ultra-broadband absorption from 5.3 to 18 GHz (a relative bandwidth of as high as 109%) is realized numerically and experimentally. In addition, simulated results demonstrate the proposed metamaterial exhibits good incident angle and polarization tolerance, which also are significant criteria for practical applications. By investigating the working principle with theoretical calculation and numerical simulation, it can be found that merging of multiple resonance modes encompassing quarter-wavelength interference cancellation, spoof surface plasmon polariton mode, dielectric resonance mode and grating mode is responsible for a remarkable ultra-broadband absorption. Analysis of respective contribution of material and structure indicates that either of them plays an indispensable role in activating different resonance modes, and symphony of material and structure is essential to afford desirable target performance. The material-structure integrated design philosophy highlights the superiority of coupling material and structure and provides an effective comprehensive optimization strategy for dielectric metamaterials.

## 1. Introduction

Microwave absorbers, aimed to trap and dissipate the microwave radiations, have been attracting significant attention for their widespread applications in electromagnetic compatibility, wireless communication transmission, stealth technology and so on. Many efforts in microwave absorbing materials have been devoted to the elaborate component and microstructure design of composites[1-4], with a purpose to achieve good impedance matching and electromagnetic (EM) energy attenuation, which capitalizes on the advantages of access to largely controllable effective EM parameters; however achievement of ultra-broadband absorption with low thickness is an enormous challenge to address owing to the single EM absorbing mechanism of quarter-wavelength interference cancellation[5, 6] or intrinsic resonances of magnetic materials, e.g. natural resonance and eddy current loss[7].

Such bottleneck issue is likely to be resolved with the emergence of metamaterials[8-10], which are artificial electromagnetic media consisting of periodic "meta-atoms" with the exotic EM properties not readily available in nature such as negative refractive index[11] and invisibility cloaking[12] etc. Metamaterial design provides a new paradigm to tailor independent electric and magnetic responses to the incident radiation, which can be an alternative philosophy for microwave absorption. Many studies on metamaterial absorbers are mainly focused on the design and optimization of metallic resonant structures[8, 13, 14], however, inevitably suffer from narrow absorption bandwidth limitation due to the nature of strong resonance characteristics. A variety of methods have been developed to broaden the absorption bandwidth, such as integration of multiple unit cells[15, 16], vertical stacked multilayer structures[17-19], and loading lumped component[20, 21]. However, these attempts increase the design and fabrication complexity and compromise flexibility accordingly. Furthermore, the issues of polarization sensitivity and incident angle dependence arising from the anisotropic unit structure may exist[22], which seriously hamper their applications in EM wave absorption.

Recently, all-dielectric metamaterials[23] have exhibited the possibility of achieving perfect absorption with resonance mechanisms different from their metallic counterparts, and offer a simpler and more versatile route for fabrication of isotropic metamaterial absorbers[24-26]; therefore, they have stimulated effusive interests. Many studies have been carried out to explore the microwave absorption of dielectric metamaterials, for example, isotropic Mie resonance-based metamaterial with cubic ceramic material[27], a series of periodic water-based metamaterials with different geometries such as droplet[28] and fishing net[29], and periodic magnetic elements with droplet shape[30] etc. These researches imply that large dielectric loss in dielectric metamaterial usually inclines to achieve high absorption[28, 30-32]. As with other materials, two essential elements of dielectric metamaterials, i.e., structure and material, allow for a large degree of freedom in tailoring performance. However, individual functionality and collaboration of material and structure in metamaterial have not been thoroughly studied and expounded.

Historically, the strategy of Materials and Structures Integrated Design (MSID) was introduced to bridge the materials engineering and structural engineering for the purpose of more effectively and flexibly achieving target structural performance[33], and is still translating, in one way or another, from a concept to a practical methodology that contributes to the high-performance/multifunctional goals with both material and structure scales optimization[34-37]. Herein, we customize MSID methodology to design ultra-broadband dielectric metamaterials and analyze the respective role of material and structure in dielectric metamaterial absorber in detail for the first time.

With the guiding principle of MSID philosophy, genetic algorithm program combined with simulation platform is adopted here to synchronously optimize the material and structure parameters of dielectric microwave metamaterial absorber (DMMA). A new DMMA with hierarchical multilayer structure is proposed here, and an ultra-broad bandwidth over the range of 5.3 - 18 GHz with absorptivity higher than 90% can be achieved and experimentally validated. Additionally, the polarization sensitivity and angular tolerance are discussed, indicating that the designed absorber can maintain high absorption under wide angles of incident wave of transverse magnetic (TM) and transverse electric (TE) mode. Furthermore, the underlying absorption mechanism is investigated via full-wave simulation and theoretical calculation, clearly demonstrating that multiple resonances including quarter-wavelength interference cancellation, spoof surface plasmon polarition (SSPP) mode, dielectric resonance mode and grating mode explain the excellent ultra-broadband absorption well. The influence of material and structure factors on absorption is discussed in detail, revealing the synergistic effects of material and structure behind the obtained excellent overall absorbing performance.

## 2. Material-structure integrated design and optimization

Carbon black/Polylactic acid (CB/PLA) composites are studied here as a model material typical of lightweight and low-cost. CB particle is a traditional and popular microwave absorber filler due to its significant frequency dispersion and EM energy dissipation, and PLA is a biodegradable and environment-friendly polyester with good processability. Materials with different EM parameters were achieved by varying the contents of CB in homogenous CB/PLA composites. The composite powders were hot-pressed at 140 ℃ into a standard toroidal shape (inner diameter: 3.04 mm, outer diameter: 7 mm, thickness: 2 ~ 4 mm) for the measurement of EM parameters. The scattering parameters ($S_{11}$, $S_{21}$) of composites were measured by vector network analyzer (R&S ZNB20) at the frequency range of 2 - 18 GHz with coaxial line. Then the complex effective permittivity $\varepsilon_r$ ($\varepsilon_r = \varepsilon_r' - \varepsilon_r''$) values of composites were retrieved from the S-parameters with Nicolson-Ross-Weir (NRW) method[38, 39]. Fig. 1(a) and (b) shows the complex permittivity of CB/PLA composites with 5wt.%, 7.5wt.%, 10wt.%, 15wt.% and 20wt.% content of CB. With the increase of concentration, the real part $\varepsilon_r'$ and imaginary part $\varepsilon_r''$ of permittivity show an obvious upward tendency and strong frequency dispersion; besides, the typical relaxation peak appears at the $\varepsilon_r''$ of higher content of 15wt.% and 20wt.%, revealing CB has a strong response to microwave. The morphology of CB/PLA composite (take 10wt.% CB/PLA composite for an example) is illustrated in Fig. S1, presenting a homogenous dispersion of CB in PLA matrix, which ensures the validity of the macroscopic effective permittivity of composites used in the simulation.

The proposed DMMA with multilayer structure comprises upper PLA honeycomb layer, intermediate anti-honeycomb resonant elements with CB/PLA composites (anti-honeycomb refers to the complementation of honeycomb structure) and bottom PLA supporting layer backed copper sheet, as shown in Fig. 1(e). Fig. 1(c) and (d) show the top view and side view of the unit cell of the DMMA, respectively. The upper honeycomb structure made of PLA with $\varepsilon_r = 2.1 \times (1 - j0.07)$ has a diameter of $d_{\text{honeycomb}}$, wall thickness of $t_{\text{honeycomb}}$ and height of $h_u$. The intermediate anti-honeycomb layer consists of array distribution of hexagonal prisms with a diameter of $d_{\text{anti-honeycomb}}$ and height of $h_m$, which is made of CB/PLA composites. The bottom PLA supporting layer with a height of $h_l$ is bounded to a copper sheet on the backside.

[Fig.1 here]

Obviously, material and structure are the essential elements of the proposed DMMA, or, they constitute the "genes" of DMMA, so to speak. As is well-known, Genetic Algorithm (GA) is an intelligent algorithm modeled on the principles and concepts of natural selection and evolution[22, 40, 41], which is usually used to find the global optimal solution of a multi-dimensional target function within the defined parameter space. Consequently, the GA combined with CST microwave studio (CST MWS) is implemented to optimize the material-structure integrated design of DMMA for broadband absorption. It must be mentioned that the GA integrated in CST MWS can only optimize the structure parameters and does not have the functionality of optimizing the material and structure parameters synchronously, which highlights the necessity of material-structure integrated optimization in this work. In particular, when the option of available materials is numerous, material-structure integrated optimization has absolute superiority of high efficiency and can free one from tedious and repetitive operation of structure optimization of all the materials using integrated GA.

A flowchart of material-structure integrated optimization via GA combined electromagnetic simulation platform is illuminated in Fig. 2. GA operates on a coding of parameters, instead of the parameters themselves; therefore, the material and structure parameters are encoded into binary strings called genes. There are 5 types of materials in this design, which can be represented by 3 binary bits. Moreover, according to the precision of the structure parameters limited to 0.01 mm and the pre-set constraints of structure parameters, 53 bits is sufficient for description of structure parameters in this work, hence each chromosome is formed by a string of 58 binary bits. It is worth noting that the number of bits depends on the range of the parameter values. Certain number of chromosomes are randomly produced as the initial population and each of them represents a DMMA (Individual). The broadband absorption is chosen as the fitness evaluation criterion. In the process, GA is realized in the MATLAB environment. MATLAB calls CST MWS via VBA (a scripting language) to send the model of DMMA, control the simulation and obtain the S-parameters at different frequencies. In the full-wave simulation from 2 to 18 GHz, the unit cell boundary is used in the $x$ and $y$-direction, and the incident wave is assigned to propagate from the $+z$ direction. In addition, because absorption of the proposed DMMA is not sensitive to the polarization of the incident wave owing to the symmetry of its structure, the transverse magnetic (TM) mode is considered here for simplification. The absorptivity is calculated from S-parameters, which is based on the equation of $A(f) = 1 - S_{11}^2(f)$ since there is no transmission due to the backed metal plate. The absorption bandwidth can be represented and fitness function can be defined by counting the average value of absorptivity higher than 90% at different frequencies. Fitness value returned by fitness function is utilized to evaluate the goodness of each individual. Some individuals with good fitness in the initial population are then selected as parents for participation in the reproduction process. The parents undergo crossover and mutation, thereby producing new children. These individuals are then inserted back into the initial population to replace individuals with low fitness, achieving generation replacement. The entire process is repeated until achieving the goal or reaching the iteration limit. Finally, the individual with the best fitness value is picked out, corresponding to the DMMA with the best match of material and structure. In conclusion, effective optimization of material-structure integrated design of ultra-broadband DMMA is achieved through the strategy of combination of GA and CST microwave studio.

[Fig.2 here]

## 3. Results and discussion

After several iterations of material-structure integrated optimization, the final parameters are obtained: 1) structure parameters ( $d_{\text{honeycomb}}$ =21.88 mm, $t_{\text{honeycomb}}$ =3.61 mm, $d_{\text{anti-honeycomb}}$ =16.80 mm, $h_{\text{u}}$ =1.48 mm, $h_{\text{m}}$ =4.87 mm, $h_{\text{l}}$ =1.27 mm); 2) material parameter (10wt.% CB/PLA composite). The simulated absorption spectrum of the designed DMMA with these optimized parameters is shown in Fig. 3(a). With reference to the figure, one can see that the DMMA achieves 90% absorption over the whole broadband of 5.3 - 18 GHz, and the relative bandwidth is up to 109%. In addition, three peak values can be observed at $f_1$ (6.3 GHz), $f_2$ (11.6 GHz) and $f_3$ (16.4 GHz), and their corresponding mechanisms will be discussed in detail later.

For comparison, the structure optimization of DMMA with 5wt.%, 7.5wt.%, 15wt.% and 20wt.% CB/PLA composite are carried out respectively via GA integrated in CST MWS and the results are exhibited in Fig. 3(b). It can be seen that the DMMA with 10wt.% CB/PLA composite and corresponding optimized structure exhibits the best absorption performance of all studied cases (broadest bandwidth and better absorption strength). Also, it has three absorption peaks, while others have two absorption peaks, indicating that ultra-broadband absorption entails a certain scale of dielectric loss, but not the larger the better and the structure contributes significantly as well. Overall, both material and structure are essential for the excellent absorption performance of DMMA. To gain a comprehensive understanding of the underlying absorption mechanism behind the optimized DMMA, the material and structure factors are analyzed separately in the following section.

[Fig.3 here]

**3.1 Material analysis**

In order to quantify the contribution of material to the absorption of DMMA, the influence of structure including upper honeycomb and middle anti-honeycomb layer should be ruled out firstly. It is necessary to transform these periodic structures to homogeneous medium models based on the homogenization technique when the period of the structure is small compared to the wavelength. It is noteworthy that the concept of average EM parameter demonstrates effectiveness at several GHz owing to the subwavelength structure of the DMMA, but the concept loses its validity at higher frequency when the elements of DMMA is large or of the order of wavelength due to the diffraction effect; therefore, the effective parameters is more meaningful at relatively lower frequencies in this work. According to the effective medium theory (EMT) based on homogenization technique[42-44], the effective permittivity of the periodic structure can be expressed approximately by:

$$\varepsilon_{eff} = \varepsilon_2 \frac{(1+f_1)\varepsilon_1+(1-f_1)\varepsilon_2}{(1-f_1)\varepsilon_1+(1+f_1)\varepsilon_2} \quad (1)$$

where, $\varepsilon_1$ and $\varepsilon_2$ represent the permittivity and $f_1 = 1 - \frac{t^2}{p^2}$ and $f_2 = 1 - f_1$ are the fractional volumes occupied by phase 1 and 2 within a single period cell (see the honeycomb structure model in Fig. 4(d)). EMT is implemented to obtain the effective permittivity of intermediate anti-honeycomb structure with 10wt.% CB/PLA composite (plotted in Fig. 4(b)) and upper honeycomb PLA structure. Then, the whole DMMA structure is equivalent to three-layer absorber backed by a perfect conductor illustrated in Fig. 4(e) and the microwave absorption performance can be calculated through transmission line (TL) theory[45, 46] with effective permittivity. According to the transmission line theory, the absorptivity of the multi-layered absorber under the TM mode of normal incident wave can be calculated by:

$$A = 1 - \Gamma^2. \tag{2}$$

$$\Gamma = \frac{Z_1 - Z_0}{Z_1 + Z_0} \tag{3}$$

$$Z_i = Z_c \frac{Z_R + jZ_c \tan(\beta d)}{Z_c + jZ_R \tan(\beta d)} \tag{4}$$

$$\beta = k_0 \sqrt{\varepsilon_r \mu_r} \tag{5}$$

where, $A$ is the absorptivity, $\Gamma$ is the reflection coefficient at the interface of free space and absorber, $Z_i$ is the input impedance between layer $i$ and $i+1$, $Z_0 \approx 377\Omega$ is the vacuum impedance $Z_c = \frac{\beta}{\omega \varepsilon}$ is the characteristic impedance, $Z_R$ is load impedance equal to the input impedance of next layer, $\beta$ is phase constant, $d$ is layer thickness, and $k_0 = \omega\sqrt{\varepsilon_0 \mu_0}$ is vacuum wavenumber. The calculated result is shown in Fig. 4(a) with a blue line. Obviously, there is a peak value with near-unity absorption located at 5.5 GHz. The electric thickness of the three-layered absorber can be calculated based on the effective permittivity:

$$D_{\text{eff}} = \sqrt{|\varepsilon_{\text{eff1}}|}\, h_u + \sqrt{|\varepsilon_{\text{eff2}}|}\, h_m + \sqrt{|\varepsilon_3|}\, h_l \tag{6}$$

where, $\varepsilon_{\text{eff1}}$ and $\varepsilon_{\text{eff2}}$ are the effective permittivity of honeycomb and anti-honeycomb structure, $\varepsilon_3$ is permittivity of bottom PLA layer, and $h_u$, $h_m$, $h_l$ is the layer thickness, respectively. From the calculated electric thickness in Fig. 4(c), one can find the electric thickness is equal to the quarter-wavelength of incident wave at 5 GHz near the peak of 5.5 GHz, demonstrating the mechanism of quarter-wavelength interference cancellation contributes much to the strong absorption at the peak of 5.5 GHz. Furthermore, the calculated absorptivity is consistent well with the simulated one before c.a. 8 GHz, which validates the effective method is reasonable until ~8 GHz since incident wavelength is several times larger than the size of unit cell and electromagnetic wave cannot "see" the inner structure of DMMA under this condition. Yet at higher frequencies, the structure cannot be deemed invisible and will have a great effect on the absorption performance. For the reliable part of calculated result, the absorption peak of 5.5 GHz is around the first peak at 6.3 GHz of DMMA, certifying that quarter-wavelength interference cancellation plays a significant role in the strong absorption of DMMA, which also verifies the dielectric material (refer to the CB/PLA composite) contributes a great deal to the microwave absorption of DMMA owing to itself excellent absorbing capacity.

[Fig.4 here]

**3.2 Structure analysis**

To further elucidate the effect of structure factors on absorption of DMMA, the field distribution in the DMMA at resonance frequency of $f_1$ (6.3 GHz), $f_2$ (11.6 GHz) and $f_3$ (16.4 GHz) are exhibited in Fig. 5, where the functionality of each layer can be analyzed intuitively.

Firstly, observing the electric field (*E*-field) at $f_1$ in Fig. 5(a) and magnetic field (*H*-field) at $f_1$ in Fig. 5(b), one can find that the incident wave is mainly confined between the anti-honeycomb element and the metal plate (or at PLA supporting layer), which is similar to the typical phenomenon of spoof surface plasmon polarition (SSPP) mode usually excited at the interface of metal and artificial dielectric resonator[47, 48]. Additionally, the majority of power loss at $f_1$ in Fig. 5(c) is distributed at the bottom of anti-honeycomb element, also indicating SSPP mode is exited and incident wave is confined at the metal surface and absorbed by anti-honeycomb element made of 10wt.% CB/PLA composite with good EM energy attenuation. Besides, there exists some power loss at the top of anti-honeycomb unit cell, which may attribute to the quarter-wavelength interference cancellation. DMMA without metal plate is simulated for comparison as shown in Fig. 6(b). The first resonance peak disappears completely, proving that SSPP mode cannot be excited without metal plate and hence it is rational to attribute the first resonance to SSPP mode. In addition, the simulated result of DMMA without bottom PLA layer (plotted in Fig. 6(c)) shows that the first peak intensity decreases dramatically, suggesting that the PLA supporting layer has a crucial effect on the SSPP mode. The above analysis of both material and structure factor reveal that hybrid mechanism of quarter-wavelength interference cancellation and resonance of SSPP mode are responsible for the strong absorption pear at $f_1$.

[Fig.5 here]

Next, the underlying principle of the second peak at $f_2$ is analyzed. From the *E*-field and *H*-field distribution at $f_2$ shown in Fig. 5(d) and (e), it can be well-observed that one part of electromagnetic field distributes between anti-honeycomb element and metal plate, which is similar to that at $f_1$ with less intensity. Another part of *H*-field distributes at the top of anti-honeycomb unit cell which exhibits a magnetic-dipole-type-like facture[24, 49], while a couple of half loop can be observed in the *E*-field. The field distribution indicates that the hybrid mode of SSPP mode and magnetic dipole mode of dielectric resonance is excited at $f_2$. In addition, according to Fig. 5(f), the majority of power loss is distributed at the top of anti-honeycomb unit cell and some at the bottom, suggesting that the dielectric resonance is dominated here. To understand the effect of periodic anti-honeycomb elements, simulation of the absorber with 10wt.% CB/PLA composite plate of the same height replacing the anti-honeycomb structure in DMMA is carried out, and the result is shown in Fig. 6(d). The average absorptivity is around 70% over the operating band, which is not sufficient for microwave absorption application. In addition, two local absorption peaks can be observed at about 2.5 GHz and 10.1 GHz. Based on the transmission line theory and effective medium theory, the calculated absorptivity of the absorber with dielectric plate is in good agreement with the simulated one shown in Fig. S2(a). Its effective electric thickness shown in Fig. S2(b) is equal to the quarter wavelength of incident wave at 2.6 GHz near the first peak of 2.5 GHz and the three-quarter wavelength at 10.0 GHz around the second peak of 10.1 GHz, indicating that both absorption peaks of the absorber with dielectric plate are the result of interference cancellation. Therefore, these new resonances in the proposed DMMA will not be excited if there are no periodic anti-honeycomb elements.

Finally, the mechanism of the third peak at $f_3$ is investigated. From the *E*-filed distribution shown in Fig. 5(g), *H*-field in Fig. 5(h) and power loss density in Fig. 5(i), one can see that *E*-field is mainly distributed at the air gap between the adjacent anti-honeycomb elements while *H*-filed is mainly on the top of anti-honeycomb elements, thus majority of the EM energy is absorbed by the top part of anti-honeycomb elements. The representative phenomenon of field distribution indicates that the grating mode due to grating diffraction effect is excited[50, 51]. The diameter of anti-honeycomb element is 16.80 mm which is close to the wavelength of 18.29 mm corresponding to $f_3$. When the periodic elements are at the scale of the wavelength, they start to scatter radiation. As such, in this situation, effective medium theory fails so that the array of elements should no longer be treated as an effective dielectric layer, but instead as a 2D grating. For in-depth understanding of the grating mode, the performance of DMMA structure made of PLA with $\varepsilon_r = 2.1 \times (1 - j0.07)$ (called all-PLA DMMA) is simulated and absorption spectrum is exhibited in Fig. 6(e). Obviously, the absorption peak arising from grating mode still exists and the other two peaks disappear, indicating that the grating mode is dominantly attributed to the appropriate structure parameters but affected very slightly by material parameters, whereas the material properties have a determining impact on the first and second absorption mechanisms. In addition, this characteristic feature of grating mode also explains why the third absorption peak at the operating frequency band only occurs to the optimized DMMA with 10wt.% CB/PLA composite but not with structure parameters corresponding to other contents: grating mode can be excited only with certain structure parameters. DMMA without upper PLA honeycomb structure is also simulated here to evaluate the effect of upper layer. As illustrated in Fig. 6(f), the first and second absorption peaks of DMMA without upper layer are almost equal to that of DMMA, while the third peak shifts to higher frequency and corresponding absorptivity shows a decreasing trend. Thus it can be concluded that the upper layer can affect the grating mode and hence the absorption performance.

[Fig.6 here]

Based on the above elucidation of the physical origin of these three absorption peaks, we can conclude that an ultra-broadband absorption of DMMA can be obtained by merging different mechanisms of quarter-wavelength interference cancellation, SSPP mode, dielectric resonance mode and grating mode via material-structure integrated design and optimization. The analysis of material and structure parameters show that both of the factors affect these resonances significantly and then absorption performance. Therefore, appropriate structural design and matching material selection is essential to acquire ultra-broadband absorption.

[Fig.7 here]

To adequately evaluate the absorption performance of the proposed DMMA under different polarizations and angles of incident wave, the performance for oblique incident wave of TE and TM mode with angles varying form $0°$ to $50°$ are simulated and results are plotted in Fig. 7. It can be observed from the absorption spectra for TM mode in Fig. 7(a) that the absorptivity remains higher than 90% over the whole operating frequency when the incident angles are below $40°$. When the angle reaches $50°$, the absorptivity is higher than 90% in majority of the frequency band and higher than 85% over the rest of the band. As for the TE mode, the absorptivity decreases at the lower frequency domain with the increasing of incident angle while remains higher than 90% at the higher frequency domain, and the absorptivity is higher than 80% over the whole operating band for angles below $40°$. The significant difference between the TM mode and TE mode can be understood by the magnetic dipole mode of dielectric resonance at the second absorption peak, which is sensitive to the alternating incident magnetic field. For the TE mode, when the incident angle increases, the effective magnetic field illuminated on the DMMA is reduced (see the inset of Fig. 7(b)). Consequently, the dielectric mode is weakened and thus the absorptivity drops significantly at the lower frequency band. In summary, the proposed DMMA exhibits good absorption performance over a broad frequency band with a wide range of incident angles.

## 4. Experimental verification

To verify the proposed design, a prototype of the DMMA was manufactured and measured. Schematic diagram of the whole preparation process is shown in Fig. S3. The 10wt.% CB/PLA composite was prepared by firstly mechanical mixing of raw materials of CB and PLA particles, and then extruding the mixture into filaments by twin screw extruder in order to ensure a homogeneous dispersion of CB particles in PLA matrix. The anti-honeycomb elements were fabricated by the hot-pressing approach within a metal mold to control the shape and size. The upper and bottom PLA layer were 3D printed. The prototype was manufactured by assembling upper layer, anti-honeycomb elements, bottom layer and a copper foil with a thickness of 0.1 mm. The photograph of the fabricated prototype with the size of 260 mm × 260 mm is presented in the Fig. 8. The weight density of the DMMA is about 0.5 g/cm², conforming well to the lightweight requirement.

The reflection coefficient of the DMMA from 2 to 18 GHz was experimentally measured by arch test system shown in Fig. S4, consisting of a vector network analyzer (KEYSIGHT E5063A) and two broadband horn antennas on arch frame. Besides, to obtain more accurate reflection of the absorber, time-domain-gating technique[52] was implemented to minimize systematic errors caused by multi-path reflections within the fixture hardware. The reflection coefficient of a metal plate with the same size as the sample was firstly measured as the reference. When the front face of the metal plate is at the same position as that of the material specimen, no phase correction is needed. It follows that the calibrated reflectivity of the sample ($S_{11}{}^{cal}$) can be expressed by the ratio of the specimen reflection ($S_{11}{}^{sample}$) to the metal plate reflection ($R_{11}{}^{short}$):

$$S_{11}{}^{cal} = \frac{S_{11}{}^{sample}}{R_{11}{}^{short}} \tag{7}$$

Absorptivity of the absorber is calculated by $A(f) = 1 - (S_{11}{}^{cal})^2$, and the result is shown in Fig. 8. It can be obviously seen that the measured absorption of the DMMA is still higher than 90% at the operating band with three local absorption peaks, which shows a good agreement with the simulated result, confirming the validity of material-structure integrated design and optimization for absorption of DMMA. There are some slight differences between simulated and measured results, which may be caused by fabrication imperfection, measurement error, and the disagreement between simulation and measurement such as idealized assumption of infinite sample mode and plane wave in simulation while sample is finite and incident wave can be approximately regarded as plane wave in measurement. Compared with the previously reported all-dielectric microwave metamaterial absorbers summarized in Table 1, the proposed DMMA in this work by material-structure integrated design demonstrates a superior overall absorption performance against the golden rule of 'broadband, strong absorption, thinness and lightweight'.

[Fig. 8 here]

Table 1.Comparison among different types of DMMA

| Reference | Type of DMMA | -10dB bandwidth (GHz) | Thickness (mm) |
| --- | --- | --- | --- |
| 30 | periodic magnetic element | 6/56% (7.8-13.8) | 4.8 |
| 53 | structured water layer | 17.6/85% (12-29.6) | 5.8 |
| 48 | cylindrical water resonator | 18.7/125% (5.6-24.2) | 5.6 |
| 54 | water-tube structure | 3.5/56% (4.5-8) | 14.5 |
| 55 | resistive square loop | 11.5/102% (5.5-17.0) | 5.5 |
| 49 | cylindrical carbon-based polymer | 8.1/102% (3.9-12) | 9.37 |
| This work | anti-honeycomb element | 12.7/109% (5.3-18) | 7.6 |

## 5. Conclusion

In summary, a dielectric metamaterial absorber with an ultra-broadband microwave absorption performance is proposed here via material-structure integrated design methodology enabled by the method of combining genetic algorithm with simulation platform. Excellent performance of DMMA requires the best matching of material and structure, which reveals that material-structure integrated design is quite instrumental for broadband absorption. As a universal method, it can be used for different EM dielectric metamaterials. Both theoretical calculation and numerical simulation are implemented to investigate the underlying mechanisms of broadband absorption and to elucidate each contribution of material and structure factors. The obtained results clearly demonstrate that merging of multiple resonances including quarter-wavelength interference cancellation, SSPP mode, dielectric resonance mode and grating mode can afford the ultra-broadband absorption. Therein, both material and structure recipes play significant roles in microwave absorption of DMMA.

Material-structure integrated design proves to be a promising strategy for excellent microwave absorber, and it has potential to further broaden the operating frequency band with more diverse materials and structure designs in the future work. Furthermore, many excellent intelligent algorithms such as machine learning and artificial neural network can be combined with simulation for high-efficiency targeted design or multi-objective optimization.

## Declaration of interest

The authors declare that they have no known competing financial interests or personal relationships that could have appeared to influence the work reported in this paper.

## Corresponding Author
***Faxiang Qin**

Tel/Fax: +86 0571 87953261. E-mail address: faxiangqin@zju.edu.cn.

## Supporting Information
Supporting information is available

## Author Contributions
**Mengyue Peng**: Investigation, Data curation, Writing- Original draft preparation. **Faxiang Qin**: Conceptualization, Methodology, Supervision, Writing- Reviewing and Editing. **Liping Zhou**: Investigation. **Huijie Wei**: Formal analysis, Writing- Reviewing. **Zihao Zhu**: Writing- Reviewing. **Xiaopeng Shen**: Data measurement

## Acknowledgements
This work is supported by ZJNSF No. LR20E010001 and National Key Research and Development Program of China No. 2021YFE0100500 and Zhejiang Provincial Key Research and Development Program (2021C01004) and Chao Kuang Piu High Tech Development Fund 2020ZL012 and Aeronautical Science Foundation 2019ZF076002.

**Figure Captions**

**Figure 1.** Material and structure parameters of the DMMA. Complex permittivity of CB/PLA composites with different contents of CB: (a) real part and (b) imaginary part. (c) Top view of the unit cell. (d) Side view of the unit cell. (e)Schematic diagram of whole DMMA with multilayer layers.

**Figure 2.** Flowchart of GA combined with CST microwave studio for material-structure integrated optimization.

**Figure 3**.Simulated absorption spectra. (a) Absorptivity of the proposed DMMA. (b) Absorptivity of the optimized DMMA with 10wt.% CB/PLA composite via material-structure integrated optimization and DMMA with 5wt.%, 7.5wt.%, 15wt.% and 20wt.% contents via structure optimization using the GA integrated in CST MWS.

**Figure 4**.Contribution of material to the absorption of DMMA. (a) Absorption spectra of simulation and equivalent calculation based on TL theory combined with EMT. (b) Effective permittivity of anti-honeycomb structure with 10wt.% CB/PLA composite. (c) Electric thickness of three-layered absorber. (d) Honeycomb structure model. (e) Equivalent three-layered structure model of DMMA.

**Figure 5**.Simulated field distribution in the DMMA. (a, d, g) E-field in xoz plane at f1, f2, f3 of 6.3 GHz, 11.6 GHz, and 16.4 GHz. (b, e, h) H-field in yoz plane at f1, f2, f3. (c, f, i) Power loss density in xoz plane at f1, f2, f3, respectively.

**Figure 6**.Simulated absorption spectra for analyzing structure factor; (a) the proposed DMMA; (b) DMMA without metal plate; (c) DMMA without bottom PLA layer; (d) the absorber with dielectric plate; (e) all-PLA DMMA; (f) DMMA without upper PLA layer.

**Figure 7.**Simulated absorption spectra of the proposed DMMA under different angles of incident wave; (a) TM mode; (b) TE mode.

**Figure 8**. Simulated and measured absorption spectra of the proposed DMMA and inset is the photo of prototype of the DMMA.

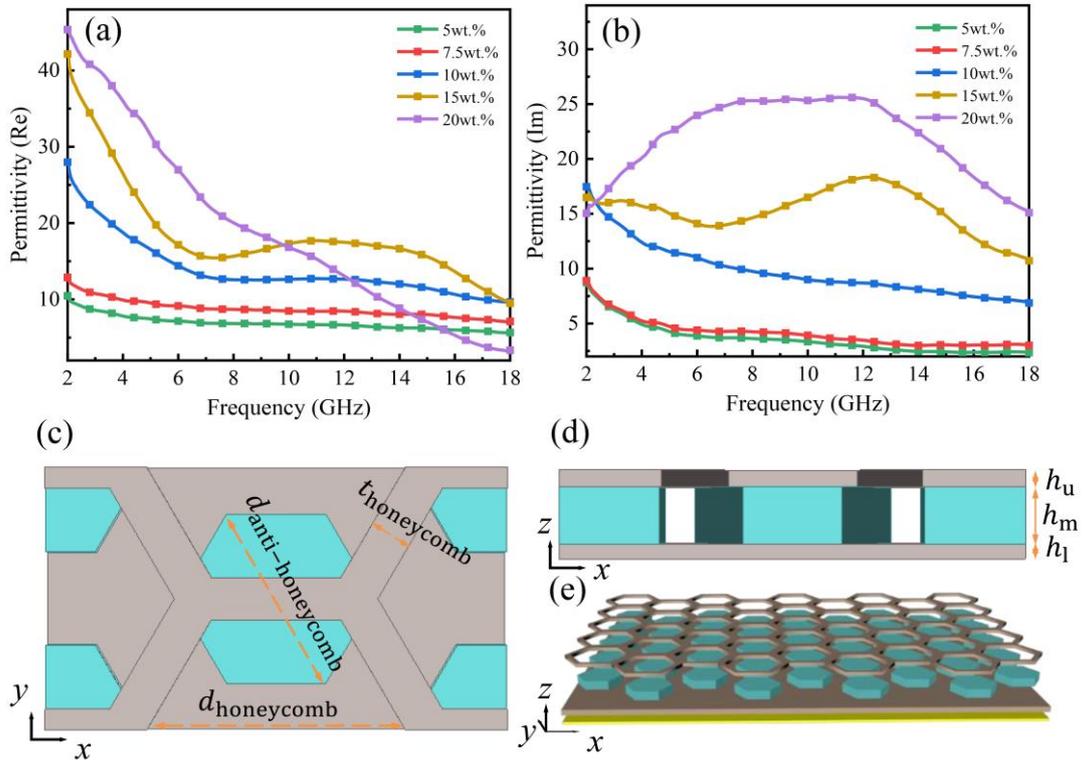

Figure 1. Material and structure parameters of the DMMA. Complex permittivity of CB/PLA composites with different contents of CB: (a) real part and (b) imaginary part. (c) Top view of the unit cell. (d) Side view of the unit cell. (e) Schematic diagram of whole DMMA with multilayer layers.

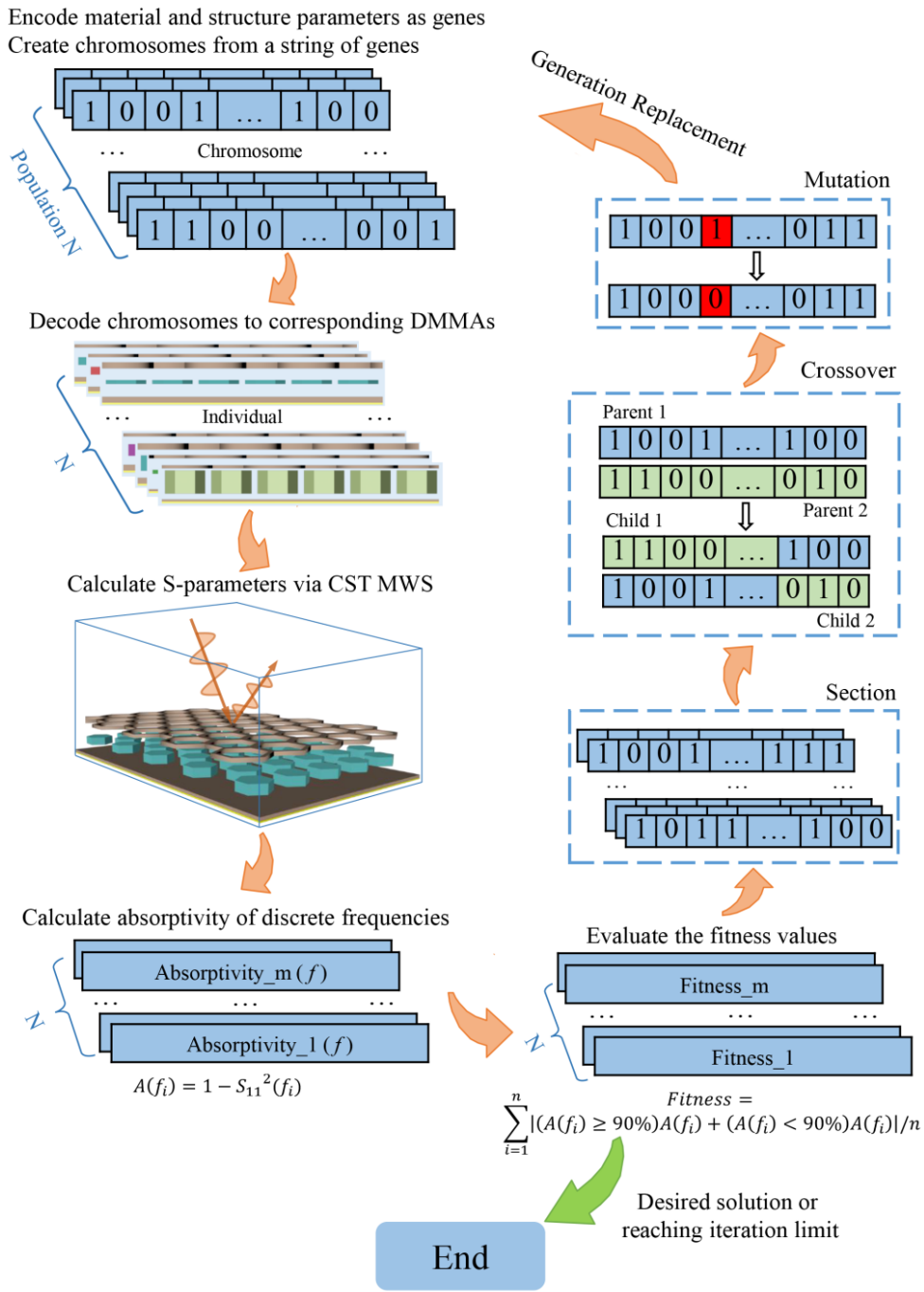

Figure 2. Flowchart of GA combined with CST microwave studio for material-structure integrated optimization.

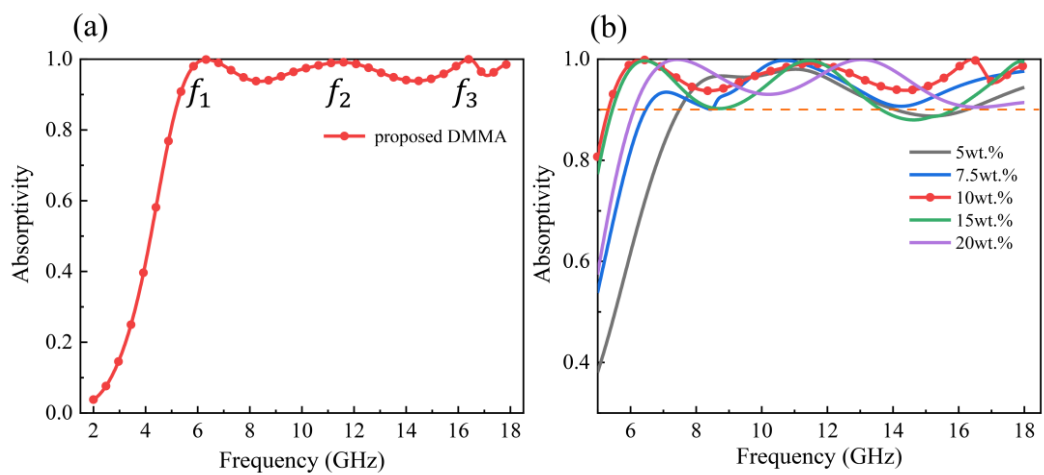

Figure 3. Simulated absorption spectra. (a) Absorptivity of the proposed DMMA. (b) Absorptivity of the optimized DMMA with 10wt.% CB/PLA composite via material-structure integrated optimization and DMMA with 5wt.%, 7.5wt.%, 15wt.% and 20wt.% contents via structure optimization using the GA integrated in CST MWS.

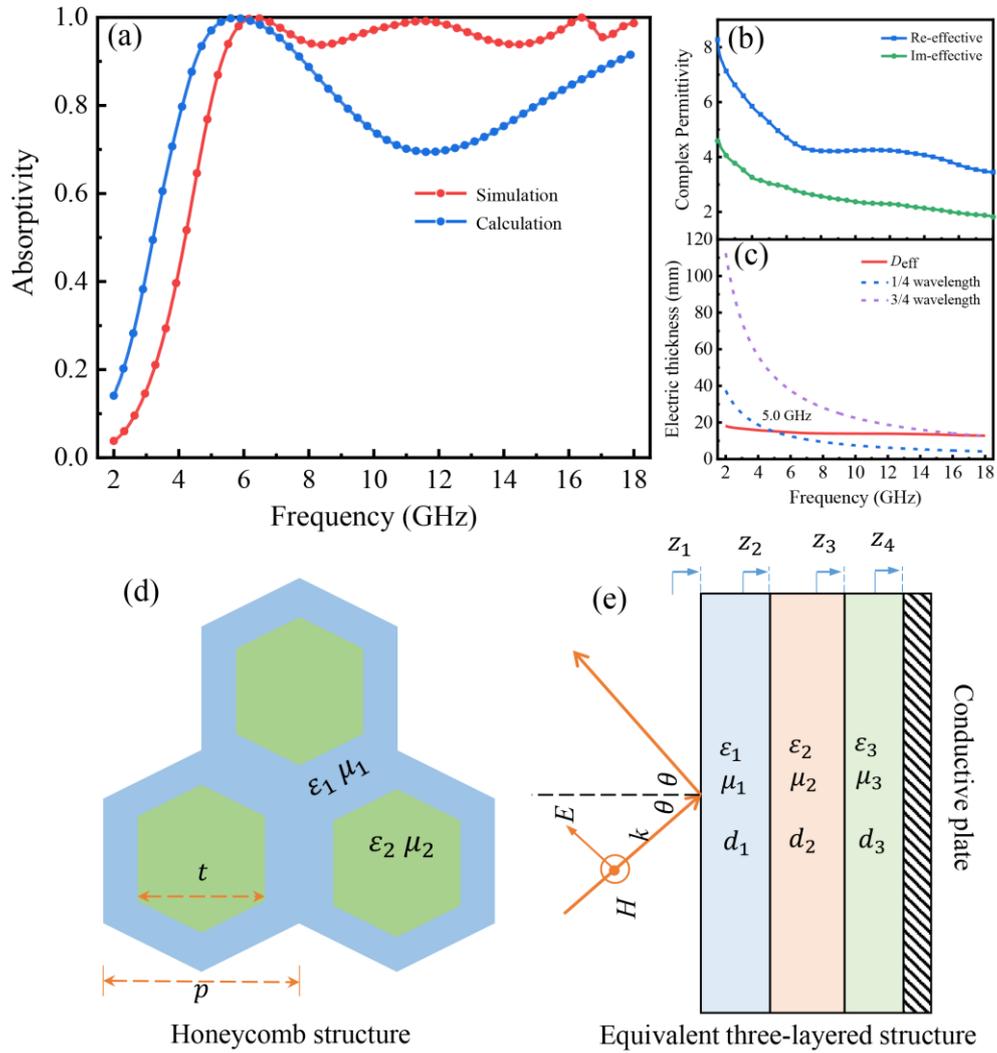

Figure 4.Contribution of material to the absorption of DMMA. (a) Absorption spectra of simulation and equivalent calculation based on TL theory combined with EMT. (b) Effective permittivity of anti-honeycomb structure with 10wt.% CB/PLA composite. (c) Electric thickness of three-layered absorber. (d) Honeycomb structure model. (e) Equivalent three-layered structure model of DMMA.

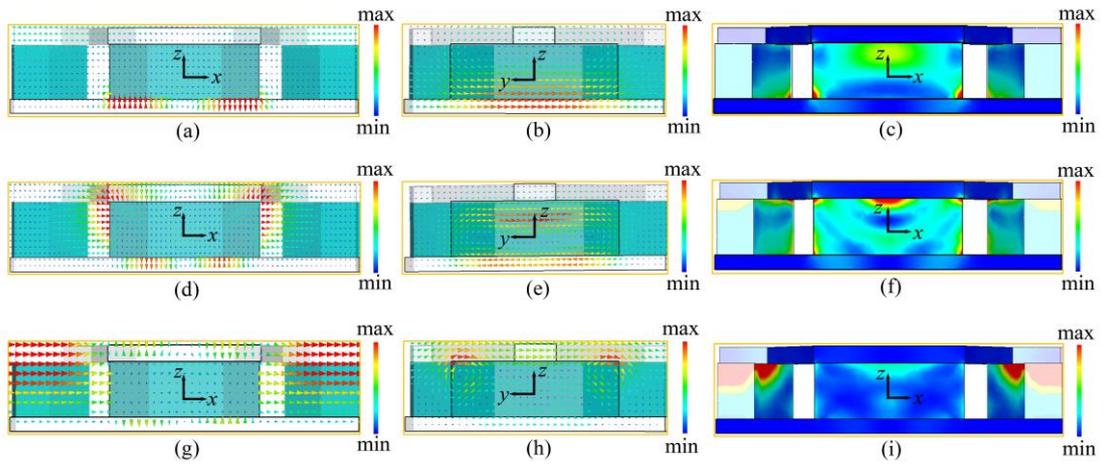

Figure 5. Simulated field distribution in the DMMA. (a, d, g) *E*-field in *xoz* plane at $f_1, f_2, f_3$ of 6.3 GHz, 11.6 GHz, and 16.4 GHz. (b, e, h) *H*-field in *yoz* plane at $f_1, f_2, f_3$. (c, f, i) Power loss density in *xoz* plane at $f_1, f_2, f_3$, respectively.

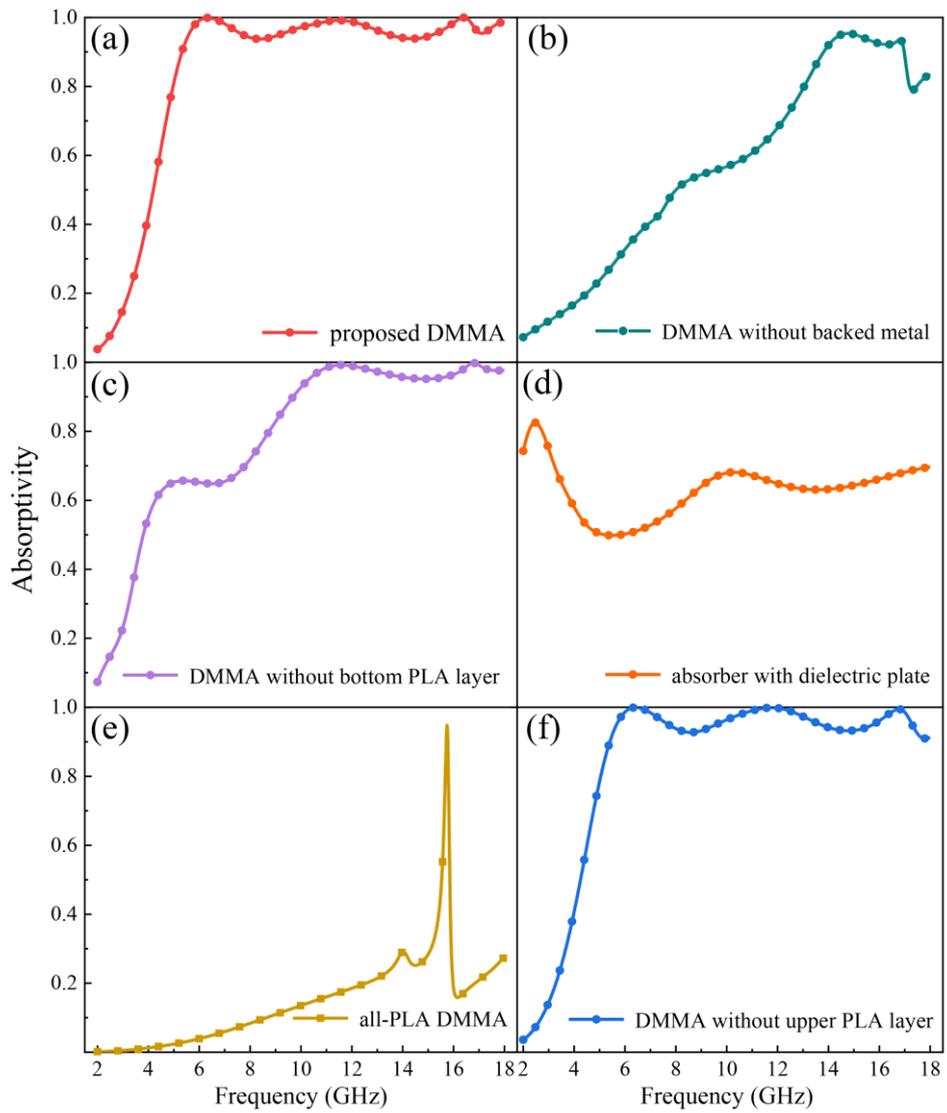

Figure 6.Simulated absorption spectra for analyzing structure factor; (a) the proposed DMMA; (b) DMMA without metal plate; (c) DMMA without bottom PLA layer; (d) the absorber with dielectric plate; (e) all-PLA DMMA; (f) DMMA without upper PLA layer.

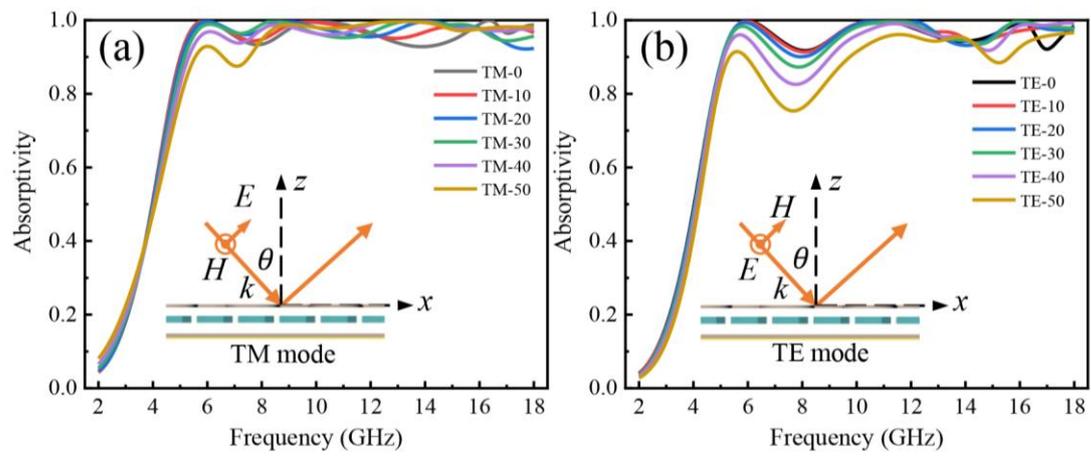

Figure 7. Simulated absorption spectra of the proposed DMMA under different angles of incident wave; (a) TM mode; (b) TE mode.

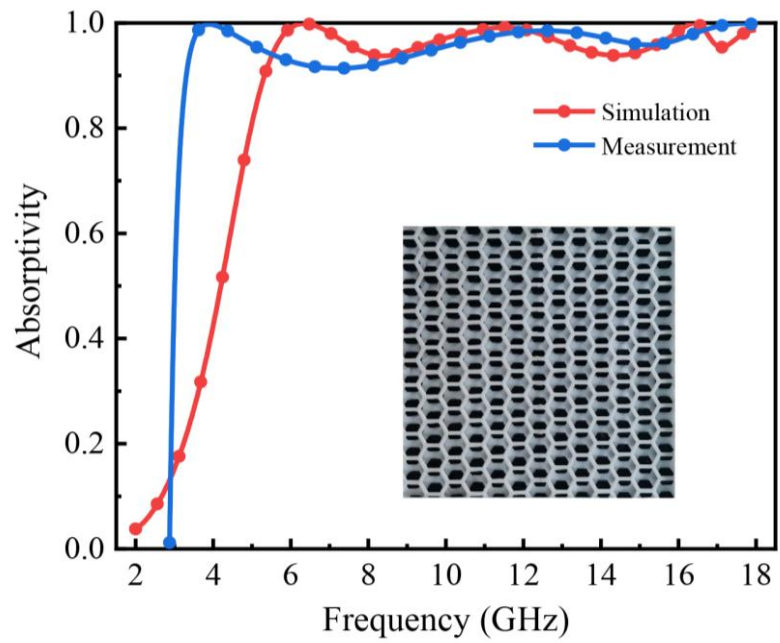

Figure 8. Simulated and measured absorption spectra of the proposed DMMA and inset is the photo of prototype of the DMMA.